# A seamless graphene spin valve based on proximity to van der Waals magnet $Cr_2Ge_2Te_6$

Haozhe Yang[1,2] *, Marco Gobbi[3,4], Franz Herling[1], Van Tuong Pham[5], Francesco Calavalle[1], Beatriz Martín-García[1,3], Albert Fert[6,7,8], Luis E. Hueso[1,3], Fèlix Casanova[1,3] *

1. CIC nanoGUNE BRTA, 20018 Donostia-San Sebastián, Basque Country, Spain
2. Fert Beijing Institute, MIIT Key Laboratory of Spintronics, School of Integrated Circuit Science and Engineering, Beihang University, Beijing 100191, China
3. IKERBASQUE, Basque Foundation for Science, 48009 Bilbao, Basque Country, Spain
4. Centro de Física de Materiales (CSIC-EHU/UPV) and Materials Physics Center (MPC), 20018 Donostia-San Sebastian, Basque Country, Spain.
5. IMEC, Kapeldreef 75, 3001 Leuven, Belgium
6. Laboratoire Albert Fert, CNRS, Thales, Université Paris-Saclay, 91767 Palaiseau, France
7. Donostia International Physics Center (DIPC), 20018 Donostia-San Sebastián, Basque Country, Spain
8. Department of Materials Physics, EHU/UPV, 20018 Donostia-San Sebastián, Basque Country, Spain

*E-mail: haozheyang.nanogune@gmail.com
       f.casanova@nanogune.eu

**Abstract:**

Pristine graphene is potentially an ideal medium to transport spin information. Proximity effects — where a neighbouring material is used to alter the properties of a material in adjacent (or proximitized) regions — can also be used in graphene to generate and detect spins by acquiring spin–orbit coupling or magnetic exchange coupling. However, the development of seamless spintronic devices that are based uniquely on proximity effects remains challenging. Here, we report a two-dimensional graphene spin valve that is enabled by proximity to the van der Waals magnet $Cr_2Ge_2Te_6$. Spin precession measurements show that graphene acquires both spin–orbit coupling and magnetic exchange coupling when interfaced with the $Cr_2Ge_2Te_6$. This leads to spin generation by both electrical spin injection and the spin Hall effect, while retaining long-distance spin transport. The simultaneous presence of spin–orbit coupling and magnetic exchange coupling also leads to a sizeable anomalous Hall effect.

## Introduction

Spintronic devices require materials that can generate and transport spin currents[1–3]. Ferromagnetic (FM) materials can, for example, generate spin currents polarized along the magnetization direction[1,2]. Alternatively, spin-orbit materials can be used for charge-spin interconversion via the spin Hall effect (SHE)[4] or the Rashba-Edelstein effect[5,6] (and their reciprocal effects). However, both the magnetic exchange coupling (MEC)[1] and the spin-orbit coupling (SOC)[4–6] reduce the spin lifetime in the materials, and thus their spin transport capabilities. As a result, different types of materials have been to be used to build functional spintronic devices, and the interfaces between them have to be carefully optimized[3,7].

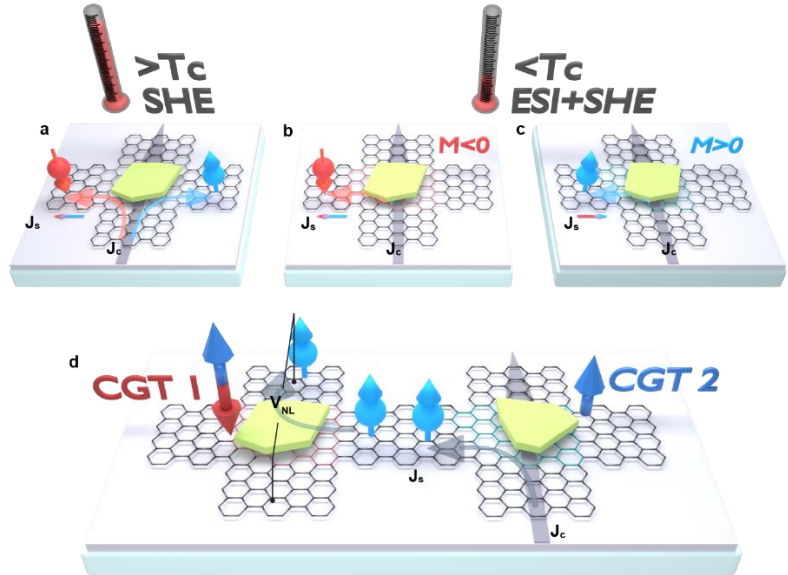

**Fig. 1 Sketch of charge-to-spin conversion and all-2D lateral spin valve in CGT-proximitized graphene**. (**a**) Spin Hall effect in spin-orbit proximitized graphene when the temperature is above $T_c$ of CGT. The spin current is generated transverse to the charge current direction with out-of-plane spin polarization. (**b, c**) Below $T_c$, CGT becomes ferromagnetic with perpendicular magnetic anisotropy, inducing magnetic proximity in graphene. The magnetization is pointing down in (**b**) and up in (**c**). The generated spin current also includes ESI from proximitized to pristine graphene. The polarization of the spin current generated by ESI changes sign when reversing the magnetization, whereas the polarization of the spin current generated by the SHE, which is not shown in (b) and (c) for clarity, remains independent from the magnetization. (**d**) An all-2D lateral spin valve with two individual proximitized graphene regions as the spin injector/detector.

Proximity effects[8,9] can, instead, be used to transform the behaviour of a given material. In such cases, a material acquires a property — superconducting, magnetic, or enhanced SOC properties, for example — of a neighbouring material in the adjacent (or proximitized) region. Emerging properties that are absent or different in the individual materials can also arise. Van der Waals heterostructures, which have short-range interfacial interaction[10], are of particular value in the design of proximitized materials[9]. Their interlayer coupling can tune the energy dispersion and spin texture of the electronic band structure[11–16], leading to emerging properties that are of potential use in the development of ultrafast and low-power electronic and spintronic devices[9,10,17].

Graphene has a low intrinsic SOC and thus exhibits a high carrier mobility[9,18,19] and long spin diffusion length[18,20]. It also provides a valuable platform to test and tailor the proximity effect, such as imprinting SOC[11–13,21] or MEC[15,22–24], which have been experimentally demonstrated individually[11,12,14,15,25–27]. Proximitized graphene (PGr) could therefore be used for spin generation in devices, while maintaining carrier mobility[18] and spin lifetime[11,28]. However, the coexistence of MEC and SOC in a two-dimensional (2D) Dirac system, which could provide a stage for studying emerging spin-dependent phenomena[29–31], has not yet been conclusively demonstrated. Furthermore, a seamless spintronic device in which all ferromagnetic metals are replaced by proximitized graphene has not yet been realized.

In this Article, we show that simultaneous magnetic and spin-orbit proximity can be achieved in graphene that is in contact with the FM van der Waals semiconductor $Cr_2Ge_2Te_6$ (CGT). We examine the generation of spin currents in a lateral spin valve based on a graphene/CGT heterostructure using non-local spin precession. Above the Curie temperature ($T_c$) of CGT, we observe a gate-tuneable SHE, which is caused by the induced SOC (Fig. 1a). Below $T_c$, along with the SHE, a MEC also emerges in the graphene, leading to electrical spin injection (ESI) that also contributes to the spin current generation (Figs. 1b and 1c).

We then construct a seamless 2D lateral spin valve that is solely composed of graphene and exhibits efficient spin generation, transport, and detection. This is achieved by adding two separate CGT flakes that induce MEC in spatially confined regions of graphene (Fig.1d). By switching the magnetization of each CGT, we observe the non-local spin valve effect, which is confirmed via a spin precession experiment. Due to the coexistence of MEC and SOC, our CGT-proximitized graphene exhibits an anomalous Hall effect (AHE).

**SOC and MEC proximity effect in CGT/graphene heterostructure**

Our CGT/graphene van der Waals heterostructures are fabricated with a deterministic dry transfer technique, and then patterned into a Hall bar geometry. CGT is a van der Waals layered semiconductor with a bandgap of ~0.7 eV (Ref. [32]) and low electrical conductivity[33,34]. Below a $T_c$ of 60-70 K (Refs. [25,35]) it becomes ferromagnetic with perpendicular magnetic anisotropy[35,36]. We verify that our CGT exhibits a p-type semiconductor behaviour, confirming that all charge transport (and, thus, spin transport) in the CGT/graphene heterostructures takes place in graphene (see Supplementary Note 1).

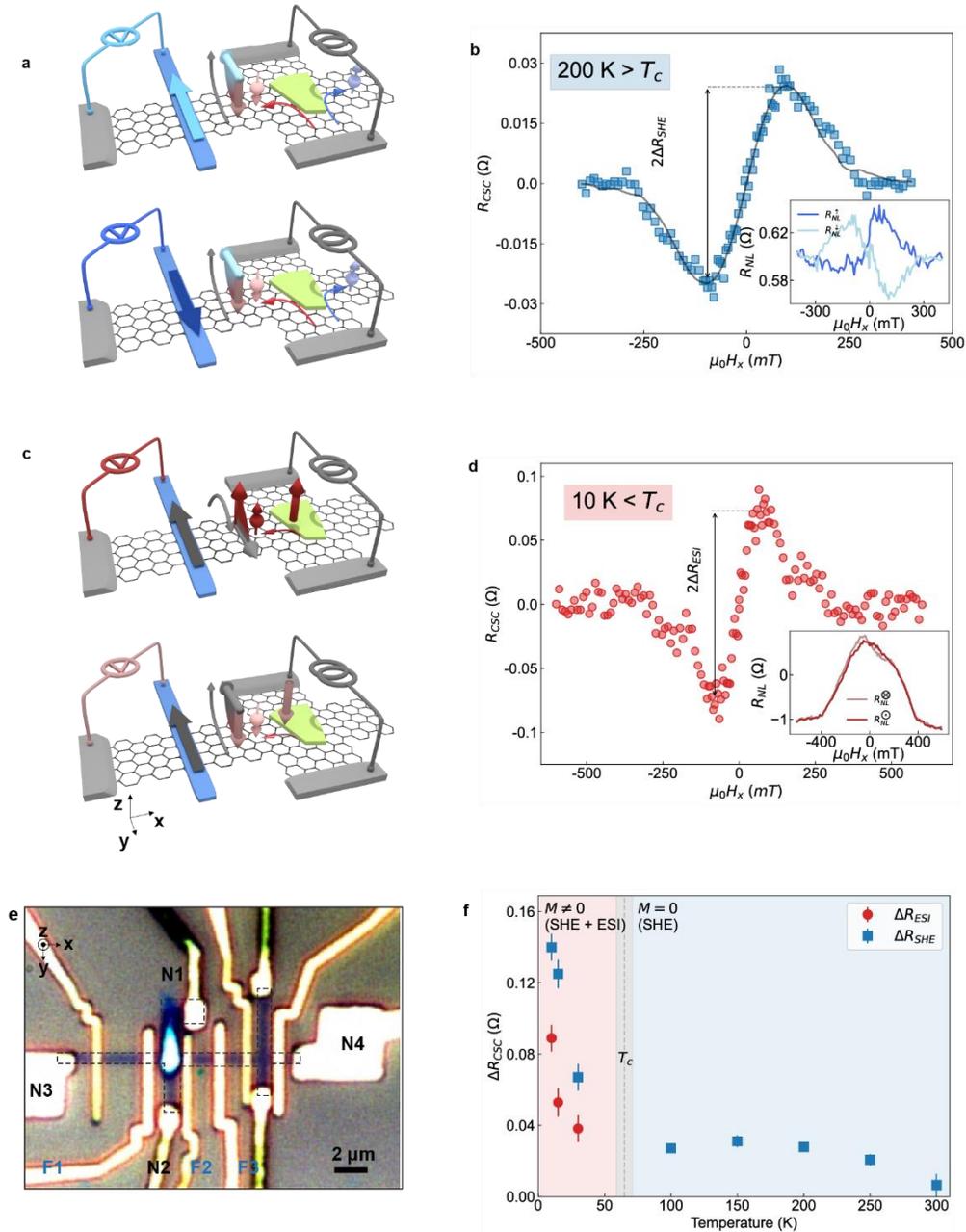

**Fig. 2 Spin Hall effect and electrical spin injection measurement for CGT-proximitized graphene.** (a) Measurement configuration for antisymmetric Hanle precession with charge-to-spin conversion. An in-plane magnetic field $\mu_0 H_x$ is applied to precess the z-polarized spins originating from the SHE of the proximitized graphene. The FM magnetization can be initialized along -y (upper panel) or +y (lower panel). (**b**) Corresponding charge-to-spin conversion raw data measured at 200 K and $V_g$ = 55 V is shown in the inset, with $R_{NL}^{\uparrow}$ for initial -y magnetization and $R_{NL}^{\downarrow}$ for initial +y magnetization. The net charge-to-spin conversion antisymmetric Hanle precession signal $R_{CSC}$ (blue squares in the main panel) is obtained from the semi-difference of $R_{NL}^{\uparrow}$ and $R_{NL}^{\downarrow}$. Fit of the data to the solution of Bloch function is shown as a gray line. The amplitude of the signal is labeled as $\Delta R_{SHE}$. (**c**) Measurement configuration for ESI. The FM electrode is always initialized along -y, while the magnetization of CGT is initialized along -z (upper panel) or +z (lower panel). (**d**) Corresponding ESI raw data measured at 10 K and $V_g$ = 55 V in shown in the inset, with $R_{NL}^{\otimes}$ for initial -z magnetization and $R_{NL}^{\odot}$ for initial magnetization +z. The net antisymmetric Hanle precession signal $R_{CSC}$ (red circles in the main panel) is obtained from the semi-difference of $R_{NL}^{\otimes}$ and $R_{NL}^{\odot}$. The amplitude of the signal is labeled as $\Delta R_{ESI}$. (**e**) Optical image of Sample 1. The dashed lines follow the edges of the graphene channel. (**f**) Temperature dependence of the antisymmetric Hanle precession amplitude originating from SHE ($\Delta R_{SHE}$, blue squares) and ESI ($\Delta R_{ESI}$, red circles). Two color-shaded regions, separated by the Curie temperature, indicate the magnetic state of CGT (red for ferromagnetic and blue for paramagnetic). Data are presented as the half-amplitude at the maximum minus the minimum of the antisymmetric curve with the error bars calculated using the standard deviation of the antisymmetric component of the Hanle precession signal within the saturation region at the highest applied fields, where the signal should be constant and allows for a consistent quantification of the noise in the measurement.

We first study the spin generation and detection of a single heterostructure based on bilayer graphene (Sample 1). Details of the device fabrication are given in *Methods*. By combining a Hall cross (with CGT on top of the cross-junction) with a FM metallic electrode made of TiO$_x$/Co (see Fig. 2e), the device enables a non-local spin precession detection scheme that isolates the charge-to-spin conversion from spurious phenomena[26,37]. Before that, we use adjacent pairs of metallic FM electrodes to calibrate the spin polarization $P_{Co}$ of the TiO$_x$/Co contact as well as the spin lifetime $\tau_s^{gr}$ and the spin diffusive constant $D_s$ of the pristine graphene (Supplementary Note 2). The charge-to-spin conversion experiment is performed by injecting a charge current ($I_c$, from 4 to 40 μA) through the Hall cross arm between non-magnetic contacts N1 and N2. The generated spin current along *x* with *z* polarization, originating from either SHE or ESI, can be detected as a non-local voltage $V_{NL}$ between magnetic contact F1 and non-magnetic contact N3. We normalized $V_{NL}$ by $I_c$ to obtain a non-local resistance $R_{NL}$. A magnetic field ($\mu_0 H_x$) is swept from zero until full saturation of the F1 contact, inducing the generated spin current with out-of-plane spins to precess in the *y-z* plane. Due to this precession, the diffusing spins gain a *y* component which is detected by F1, resulting in an antisymmetric Hanle precession curve, with a maximum and a minimum at a certain $\pm\mu_0 H_x$ value.

To disentangle the source of the charge-to-spin conversion between SHE and ESI, we perform the measurements above and below $T_c$ of CGT in order to turn off and on the magnetic proximity of PGr. The configuration for detecting SHE is shown in Fig. 2a. The magnetization direction of F1 electrode (+y or −y) can be set with an external magnetic field ($\mu_0 H_y$) along their easy axis. When switching F1, the antisymmetric Hanle curve $R_{NL}^{\downarrow}$ is reversed with respect $R_{NL}^{\uparrow}$ because the detector senses the opposite y-spin component. One representative result measured at 200 K and $V_g$ = 55 V is shown in the inset of Fig. 2b. The pure charge-to-spin precession signal due to the SHE, plotted with blue squares in Fig. 2b, has been obtained from $R_{CSC} = (R_{NL}^{\uparrow} - R_{NL}^{\downarrow})/2$, followed by an antisymmetrization of the obtained curve to the magnetic field. This process removes any initial-magnetization independent components, such as local magnetoresistance, ordinary Hall effect[37] and conventional Edelstein effect[26,38], as well as the possible contribution of unconventional charge-to-spin conversion with spins polarized along y, which has been observed in some van der Waals heterostructures[38–42]. We fitted the antisymmetric Hanle curve (gray line in Fig. 2b) with a numerical model, after fixing $P_{Co}$=3.8%, $\tau_s^{gr}$=65 ps and $D_s$=0.0069 $m^2/s$ obtained from the previous fit in Supplementary Note 2. In order to decrease the number of fitting parameters, we assume $D_s$ to be the same in the pristine and proximitized graphene regions. We obtain a spin Hall angle of $\theta_{SH} = 0.8 \pm 0.6$ %, and a spin diffusion length $\lambda_s^{PGr}$= 550 nm ± 940 nm at 200 K and $V_g$ = 55 V, with the fitting process detailed in Supplementary Note 10.

Next, we turn on the magnetism by decreasing the temperature below $T_c$ of CGT. The polarization of the generated spins due to spin-dependent conductivity of the magnetic proximitized graphene should follow the magnetization of CGT (±z), whereas the spin polarization originating from the SHE is independent of the CGT initial magnetization direction. Another z-polarized spin current source when magnetism is activated is the spin anomalous Hall effect[43–46], but it is a time-reversal-even effect and thus experimentally indistinguishable from the conventional SHE-generated spin current[47]. In order to disentangle the spin signal originated by ESI from the one induced by the SHE, we use the measurement configuration shown in Fig. 2c. The magnetization direction of F1 electrode is fixed to be −y, and the magnetization of CGT along its easy axis +z, obtaining the non-local resistance $R_{NL}^{\odot}$. When reversing the initial CGT magnetization along −z, $R_{NL}^{\otimes}$ originating from ESI is reversed because the spin current reverses its polarization. To perform this experiment, we first initialize the CGT magnetization with a large in-plane magnetic field with a specific polarity and then release the field, thereby orienting the CGT magnetization in a preferred direction. The magnetization of the Co is subsequently initialized along −y by applying a small in-plane y-field. The stable magnetization direction of Co at higher magnetic fields is evidenced by the symmetric Hanle precession shown in Supplementary Note 2 (Fig. S2b). Finally, we sweep the magnetic field ($B_x$) from zero until the full saturation of the F1 and CGT, which we repeat for the 4 possible configurations of magnetization of CGT and polarity of the magnetic field ($\otimes \rightarrow$, $\otimes \leftarrow$, $\odot \rightarrow$, and $\odot \leftarrow$). No switching of the CGT is observed during the preparation of Co, as this would give the same $R_{NL}$ curve during the hard-axis scan of the field with a given polarity, which is clearly not detected. An exemplary result, measured at 10 K and $V_g$ = 55 V, is plotted in the inset of Fig. 2d. By subtracting the $R_{NL}$ curves between the positive ($R_{NL}^{\odot}$) and negative ($R_{NL}^{\otimes}$) alignment of the CGT magnetization, $R_{CSC} = (R_{NL}^{\otimes} - R_{NL}^{\odot})/2$, also followed by an antisymmetrization, the net ESI-induced spin precession signal remains, while removing any signal which is independent of the initial CGT magnetization, as shown in Fig. 2d. By averaging $R_{NL}^{\otimes}$ and $R_{NL}^{\odot}$, the ESI-induced spin precession signal is cancelled, and we can thus extract the SHE-induced antisymmetric spin precession component by removing the symmetric background. Such background, which shows a saturation field that corresponds to the in-plane saturation field of CGT, may arise from two possible origins: a local magnetoresistance of the magnetic graphene induced by stray current from the non-local measurement, or a magnetic multidomain state of CGT. Nonetheless, the antisymmetric signal appears well below the in-plane saturation field of CGT, indicating a spin precession origin, as discussed in detail in Supplementary Note 3. The temperature dependence of the antisymmetric Hanle precession amplitude originating from SHE ($\Delta R_{SHE}$, blue squares) and ESI ($\Delta R_{ESI}$, red circles) measured at $V_g$ = 55 V is shown in Fig. 2f. Both amplitudes increase when decreasing the temperature. The SHE amplitude increases when decreasing the temperature, as has been observed in graphene with other spin-orbit proximity systems[12,14]. The possible mechanism of the observed SHE is discussed in Supplementary Note 11. Below $T_c$, both amplitudes exhibit a more pronounced increase. The enhancement of $\Delta R_{SHE}$ could arise from the additional contribution of the spin anomalous Hall effect when the magnetism is turned on. The enhancement of $\Delta R_{ESI}$ is expected because in a ferromagnet, the spin polarization increases with decreasing temperature below $T_c$[48]. The SHE would be contingent on the magnetization of CGT if the spin anomalous Hall effect is considered, since this contribution does correlate, in first approximation, with the polarization of the ferromagnet.

To generate a spin current by ESI, one needs a charge current to flow from a ferromagnet to a non-magnet, where the spin accumulation is created at the interface[20]. Such spin accumulation is expected to cancel out in a perfectly symmetric Hall cross. In our device, however, the geometry of the CGT flake is slightly asymmetric, leading to a net spin accumulation. This spin accumulation due to inhomogeneous ESI has also been observed in an all-metallic lateral spin valve[49]. Here, we use a 3D finite element simulation method to extract the ESI efficiency in our particular device geometry, shown in Supplementary Note 4. The fitting result gives a spin polarization $P_{PGr}$ = 14.5% ± 9.6 %, and a spin diffusion length $\lambda_s^{PGr}$ = 250 nm ± 890 nm of the proximitized graphene at 10 K and $V_g$ = 55 V. This ESI is superimposed to the SHE, for which we estimate $\theta_{SH}$ = 2.4 ± 1.6 %. As expected, the charge-to-spin conversion is tuneable with a back gate, showing a maximum around the charge neutrality point of the PGr (see Supplementary Note 5).

**All-2D lateral spin valve**

After studying the charge-to-spin conversion behaviour of the CGT/graphene heterostructure, we then build an all-2D, seamless lateral spin valve, sketched in Fig. 3a. Two separated proximitized graphene regions are connected through pristine graphene. The optical image of Sample 2 is shown in Fig. 3b. Independently exfoliated CGT flakes are stamped on one exfoliated single layer graphene, then patterned into a Hall bar geometry. Since theoretical calculations show that single layer[50] and bilayer[29] graphene in proximity with CGT exhibit similar MEC energy, we expect a magnetic proximity in Sample 2 similar to Sample 1. Due to the different sizes and thicknesses (shown with AFM images in Supplementary Note 6), the CGT flakes have different coercivity, which allows them to be switched individually. A charge current $I_c$ is injected into one proximitized region (PGr1), and creates a transverse spin current flowing through the pristine graphene. This spin current is converted back into a charge current when reaching the second proximitized region (PGr2) and is detected as a non-local voltage $V_{NL}$ (again normalized into a non-local resistance, $R_{NL} = V_{NL}/I_c$). When PGr1 and PGr2 magnetizations are parallel (P) or antiparallel (AP), $R_{NL}$ exhibits a different value because of the well-known non-local spin valve effect[51], reported here for the first time in a fully proximitized system with no ferromagnetic metals. $R_{NL}$ measured from 5 to 50 K and $V_g$ = 50 V is shown in Fig. 3c by sweeping the magnetic field out of plane ($\mu_0 H_z$). We can clearly observe two $R_{NL}$ states, which correspond to the P and AP states of the two PGr. Here, the non-local spin valve signal ($\Delta R_{NL}^{LSV}$, as defined in Fig. 3c) is originated from the coexistence of ESI and (inverse) SHE, which has been discussed in Supplementary Note 9.

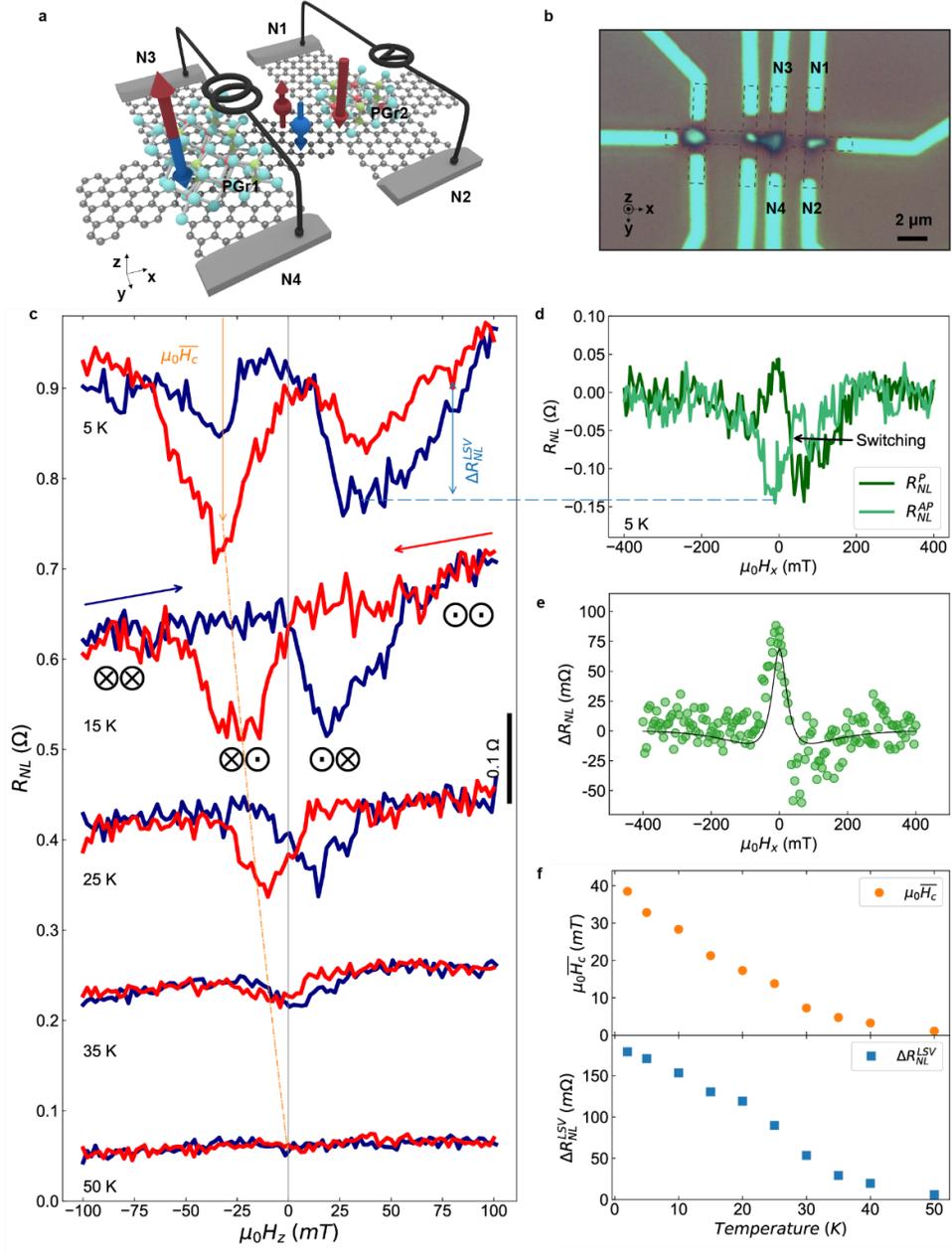

**Fig. 3 All-2D spin valve with two CGT-proximitized graphene regions.** (**a**) Schematic representation and (**b**) optical image of an all-2D lateral spin valve from Sample 2. (**c**) Non-local resistance $R_{NL}$ collected

at 5, 15, 25, 35, and 50 K with $V_g$ = 50 V, while sweeping $\mu_0 H_z$; blue/red curves indicate trace/retrace direction. The two resistance states for parallel and antiparallel configurations are labeled in the curve of 15 K. Curves are vertically offset for clarity. The orange line illustrates the decrease in switching field as the temperature increases, while the grey line represents the zero magnetic field. The non-local spin valve signal $\Delta R_{NL}^{LSV}$, defined as the difference between $R_{NL}$ at the parallel and antiparallel states, is labelled. (**d**) Hanle spin precession raw data, obtained at 5 K and $V_g$ = 50 V as a function of $\mu_0 H_x$, for parallel ($R_{NL}^P$, dark green line) and antiparallel ($R_{NL}^{AP}$, light green line) configurations. A switching event during the field sweep is labeled. (**e**) Net spin precession signal $\Delta R_{NL}$ (green circles) obtained from the semi-difference of $R_{NL}^P$ and $R_{NL}^{AP}$. Fit of the data to the solution of 1D Bloch function is shown as a gray line. (**f**) $\Delta R_{NL}^{LSV}$ and average coercive field $\mu_0 \overline{H_c}$, extracted from (c), as a function of temperature. The error bars are derived from the variance between the values obtained from the positive and negative fields of the non-local resistance.

We further confirm the spin current origin of the non-local signal with a spin precession experiment. The two PGr regions are initialized into a P or AP configuration with $\mu_0 H_z$, shown with two guided dashed lines linking the $\mu_0 H_z$-scan signal in Fig. 3c at the same condition. Then, a magnetic field ($\mu_0 H_x$) along the hard axis *x* of the PGr is swept from zero until full saturation of the PGr. At lower fields, spins diffusing into the pristine graphene channel precess in the *y-z* plane and $R_{NL}$ exhibits a symmetric Hanle precession behavior. Raw data measured at 5 K and $V_g$ = 50 V is shown in Fig. 3d. The magnetization control along the hard axis scan is notably effective, although some switching between P and AP states during the positive field scan along the hard axis occurs, as indicated. The minor switching can also be observed in Fig. 3c, as it might arise from the multidomain state of the CGT, which has been observed with other methods such as SQUID-on-tip microscopy[36], magnetic circular dichroism[52], or vector-field cryogenic magnetic force microscope[53]. The pure spin precession signal $\Delta R_{NL} = (R_{NL}^P - R_{NL}^{AP})/2$, plotted as green circles in Fig. 3e, shows the typical symmetric Hanle curve, confirming the spin transport origin of the signal and ruling out possible artifacts. We used a 3D finite element simulation method to fit the data, obtaining a spin polarization $P_{PGr}$ = 61.5 ± 9.5 % and a spin diffusion length of the pristine graphene $\lambda_s^{gr}$ = 1240 ± 170 nm, the latter in agreement with what is expected for pristine graphene (Supplementary Note 4). Since the actual spin polarisation value should be obtained from a single magnetic domain, a multidomain configuration would lead a reduced effective spin polarization. The spin polarisation here is in the same order of magnitude but larger than the one estimated from Sample 1, a difference which could be attributed to different magnetic multidomain configurations of the CGT flakes, with different shapes and thicknesses. Future studies, using optimised single-domain CGT flakes or incorporating magnetic domain microscopy and electrical transport measurements, could be employed to determine spin polarisations with greater accuracy. The temperature dependence of the all-2D lateral spin valve shows that the spin signal of the AP state is more robust at a lower temperature, as the coercive fields of the two CGT increase, and vanishes when the temperature is close to the $T_c$ of CGT. The average coercive field ($\mu_0 \overline{H_c}$) of PGr1, determined by taking the mean of the coercive fields corresponding to the positive and negative magnetic field directions, decreases when increasing the temperature, along with $\Delta R_{NL}^{LSV}$ (Fig. 3f).

**AHE enabled by the coexisting proximity effects**

Once the coexistence of SOC and MEC proximity is confirmed in the graphene/CGT van der Waals heterostructure, we also expect the emergence of AHE in graphene, which has been previously observed in some in-plane magnetic anisotropy systems[15,16,54,55]. For the all-2D lateral spin valve device, we are able to detect both the non-local spin valve effect and the AHE, see Fig. 4a. A representative AHE in Sample 2, along with the non-local spin valve effect, is shown in Supplementary Note 8. Here, we present another device (Sample 3) with a larger coercive field, providing a clearer result. The characteristic geometric and magnetic parameters of all three samples are listed in **Table 1**. The optical image of Sample 3, based on a bilayer graphene, is shown in Fig. 4b, with a distance of 3.6 μm between the two proximitized regions. Figure 4(c) plots the Hall resistance ($R_H$) measured at 30 K and $V_g$ = 30 V, where a linear background originating from the ordinary Hall effect has been subtracted, evidencing a clear hysteretic behaviour associated to the AHE (with amplitude $\Delta R_{AHE}$). In Fig. 4d, we plot the non-local resistance at the same conditions, showing clear P and AP states. The vertical dot-dashed lines show an excellent agreement of the coercive field between the AHE and the non-local spin valve measurements.

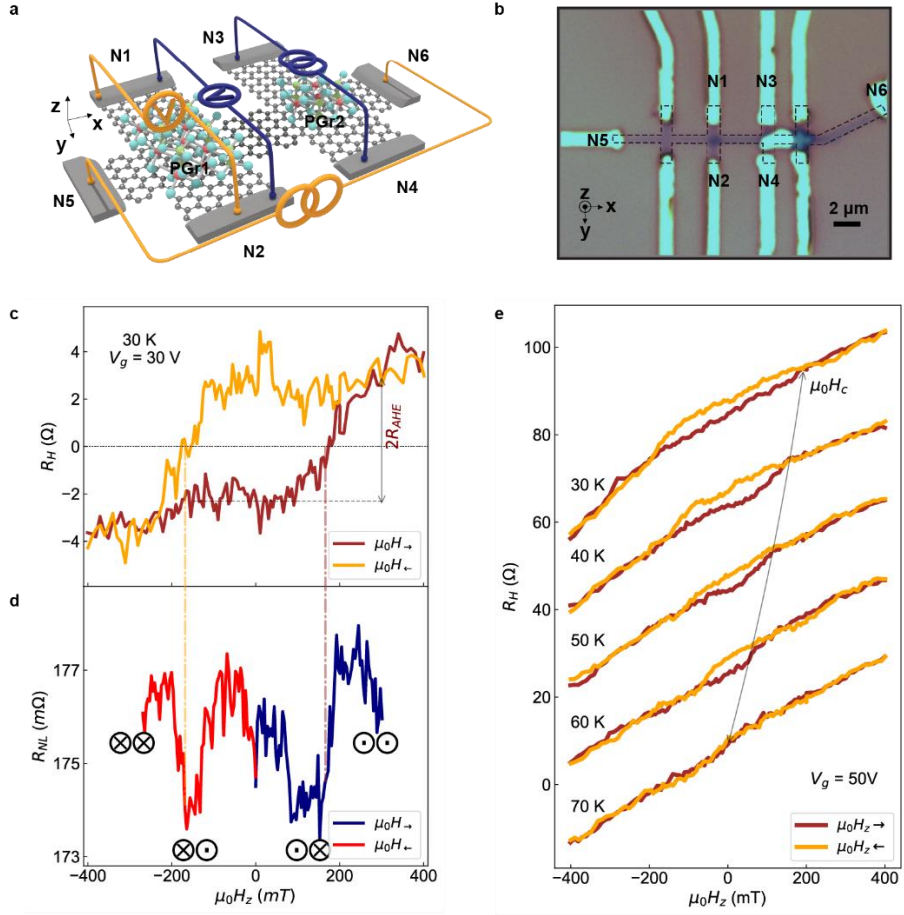

**Fig. 4 Anomalous Hall effect of CGT-proximitized graphene.** (**a**) Measurement configuration for non-local spin valve signal (blue wires) and for anomalous Hall effect (yellow wires). (**b**) Optical image of Sample 3. (**c**) Hall resistance $R_H$ (after subtracting the linear background and with the AHE amplitude labeled as $2\Delta R_{AHE}$) and (**d**) non-local resistance $R_{NL}$ (with the two states for P and AP configurations labeled) as a function of $\mu_0 H_z$ measured at 30 K and $V_g$ = 30 V. The switching field of PGr 1 in (**d**) corresponds to the coercive field in (**c**), illustrated with two dot-dashed lines. (**e**) $R_H$ measured from 30 K to 70 K at $V_g$ = 50 V, indicating the decrease of the coercive field guided with the grey lines. The curves are vertically shifted for clarity.

The temperature dependence of $R_H$ (without subtracting the linear background) is shown in Fig. 4e, measured from 30 to 70 K at $V_g$ = 50 V. An apparent out-of-plane magnetic anisotropy of PGr, with a well-defined hysteretic behaviour, is observed below $T_c$. The AHE disappears when reaching 70 K, the $T_c$ of CGT. A grey arrow (guide to the eye) illustrates the coercive field $\mu_0 H_c$, showing a decrease from 170 mT at 30 K to 50 mT at 60 K, and $\Delta R_{AHE}$ decreases from 2.6 Ω to 1.8 Ω. Although these values are far from the resistance quantum, the observed coexistence of

MEC and SOC in a 2D Dirac material makes this platform a solid candidate for quantum anomalous Hall effect[19,30,56]. In this regard, there is still much room for decreasing the disorder energy scale to become smaller than the SOC in our devices, for instance by encapsulation with hexagonal boron nitride.

Table 1 Summary of characteristic geometric and magnetic parameters of the samples used in this work.

|  | Sample 1 | Sample 2 | Sample 3 |
| --- | --- | --- | --- |
| **Graphene thickness** | Bilayer | Monolayer | Bilayer |
| **CGT flake thickness** | 50 nm | 20 nm<br>17 nm | 85 nm<br>60 nm |
| **CGT dimensions (minimum bounding rectangle)** | 6.7 μm ×1.8 μm | 0.9 μm ×1.8 μm/<br>1.5 μm ×1.6 μm | 0.5 μm ×0.7 μm/<br>1.7 μm ×2.9 μm |
| **Coercivity field $\mu_0 \overline{H_c}$ (T = 30 K)** | 45 mT | 25 mT/<br>8 mT | 170 mT/<br>94 mT |

**Conclusions**

We have shown that spin-orbit coupling and magnetic exchange coupling can be simultaneously present in graphene due to proximity to the ferromagnetic van der Waals material $Cr_2Ge_2Te_6$. With the system, we observed both the spin Hall effect and electrical spin injection. Based on this interface, we constructed an all-2D, seamless lateral spin valve that exhibits a spin valve effect signal by individually switching the two magnetic graphene regions. The spin current origin of the signal is corroborated through spin precession measurements. The magnetic multidomain state of CGT result in a variable spin injection, which could be possibly optimised by controlling the layer number and shape of the CGT flake. We also showed that the simultaneous presence of spin-orbit and magnetic proximities allows the anomalous Hall effect to be observed in the graphene/$Cr_2Ge_2Te_6$ van der Waals heterostructure. Our work provides insight into the fundamental physics of spin transport in van der Waals heterostructures and could also help in the development of spintronic applications based on proximities within a 2D Dirac system.

## Methods:

### Sample fabrication

Graphene flakes are first produced by mechanical exfoliation of natural graphite (supplied by NGS Naturgraphit GmbH) onto 300-nm-thick $SiO_2$ on doped Si substrate using Nitto tape (Nitto SPV 224P) and identified using its optical contrast in air. After selecting graphene flakes, all the samples are put into an Ar glove box with $O_2$ concertation below 1.0 ppm and $H_2O$ below 0.5 ppm. Inside the glovebox, the $Cr_2Ge_2Te_6$ crystal (supplied by HQ Graphene) is exfoliated using the Nitto tape and transferred onto a piece of polydimethylsiloxane (Gelpak PF GEL film WF 4, 17 mil.). After identifying the desired flake using an optical microscope, it is stamped on top of the graphene flake using a viscoelastic stamping tool in which a three-axis micrometer stage is used to position the flake accurately.

After the stamping, the graphene flake is then nanopatterned into Hall bars, using electron-beam lithography, and covered with a 20-nm-thick Al deposited by high-vacuum thermal evaporation (base pressure $\sim 7 \times 10^{-7}$ Torr) as a hard mask. The device is then reactive-ion etched with $Ar/O_2$ plasma. The Al hard mask is then chemically removed with a base developer solution with tetra-methyl ammonium hydroxide, producing no evident damage in CGT nor in graphene, as shown by the Raman spectra in Supplementary Note 7. The sample is annealed at 400 °C for 60 minutes in ultrahigh vacuum ($10^{-9}$ Torr) to remove fabrication residues. For Sample 1 and Sample S1, 45-nm-thick sputtered Pt was used to contact graphene. Sample 2 and Sample 3 are contacted with Ti(2 nm)/Au(55 nm) by electron-beam evaporation and lift-off in acetone. For Sample 1 and Sample S1, the magnetic $TiO_x$/Co electrodes were fabricated by e-beam lithography, deposition of 3 Å of Ti (followed by oxidation in air for 10 minutes), 35 nm of Co, and 10 nm of Au as capping layer by e-beam evaporation, and lift-off. The presence of $TiO_x$ between the Co electrode and the graphene channel leads to interface resistances between 12 and 50 k$\Omega$, measured with a conventional three-point measurement. The width of the magnetic electrodes are 150 nm or 300 nm in order to have different coercivity.

### Electrical measurements:

The measurements are performed in a Physical Property Measurement System (PPMS) by Quantum Design, using a DC reversal technique with a Keithley 2182 nanovoltmeter and a 6221 current source. The n-doped Si substrate acts as a back-gate electrode to which we apply the gate voltage with a Keithley 2636B.

**Devices characterization:**

The exact dimensions of the devices are extracted from scanning electron and atomic force microscopy images obtained after the electrical measurements, shown in Supplementary Note 7. Micro-Raman spectroscopy measurements were carried out at room temperature after the electrical characterization, by placing the samples in a Linkam® vacuum chamber ($10^{-6}$ hPa) coupled to a Renishaw® inVia Qontor Raman instrument equipped with a 50× objective (Nikon®, N.A. 0.60, WD 11 mm). We use a 532 nm laser as excitation source with a 2400 l/mm grating. The laser power was kept < 0.5 mW to avoid the damage of the samples during the measurement.

**Data and code availability**

Source data in the paper are available at https://doi.org/10.6084/m9.figshare.22815824. Any further data and codes used in this study are available from corresponding authors upon reasonable request.

**Acknowledgements**

We thank Vivek Amin for useful discussions. We acknowledge funding from the "Valleytronics" Intel Science Technology Center; from the Spanish MICIU/AEI/10.13039/501100011033 ("Maria de Maeztu" Units of Excellence Programme grant No. CEX2020-001038-M); from MICIU/AEI and ERDF/EU (Project Nos. PID2021-122511OB-I00 and PID2021-128004NB-C21); from Diputación de Gipuzkoa (Project No. 2021-CIEN-000037-01); from the European Union's Horizon 2020 research and innovation programme under the Marie Skłodowska-Curie grant agreement No 766025 (QuESTech). H.Y acknowledges support from NSFC 92164206 and 52261145694. This work was also supported by the FLAG-ERA grant MULTISPIN, via the Spanish MICIU/AEI and European Union NextGenerationEU/PRTR with grant number PCI2021-122038-2A. B.M.-G. and M.G. acknowledge support from the "Ramón y Cajal" Programme by the Spanish MICIU/AEI and European Union NextGenerationEU/PRTR (grant Nos. RYC2021-034836-I and RYC2021-031705-I, respectively). A. F. acknowledges the support of the UNIVERSIDAD DEL PAIS VASCO as distinguished researcher.

**Author contribution:**

H.Y. and F. Casanova conceived the study. H.Y. performed the sample fabrication with the help of F.H., H.Y. performed the electrical measurements with the help of F. Calavalle and F.H. B.M.-G and H.Y. performed the Raman spectroscopy measurements. V.T.P and H.Y. performed the finite element simulation. All authors contributed to the discussion of the results and their interpretation. H.Y., M.G. and F. Casanova wrote the manuscript with input from all authors.

**Competing interests**

The authors declare no competing interests.

# SUPPLEMENTARY INFORMATION

**Table of contents**

**Note 1. Electrical characterization of CGT.**

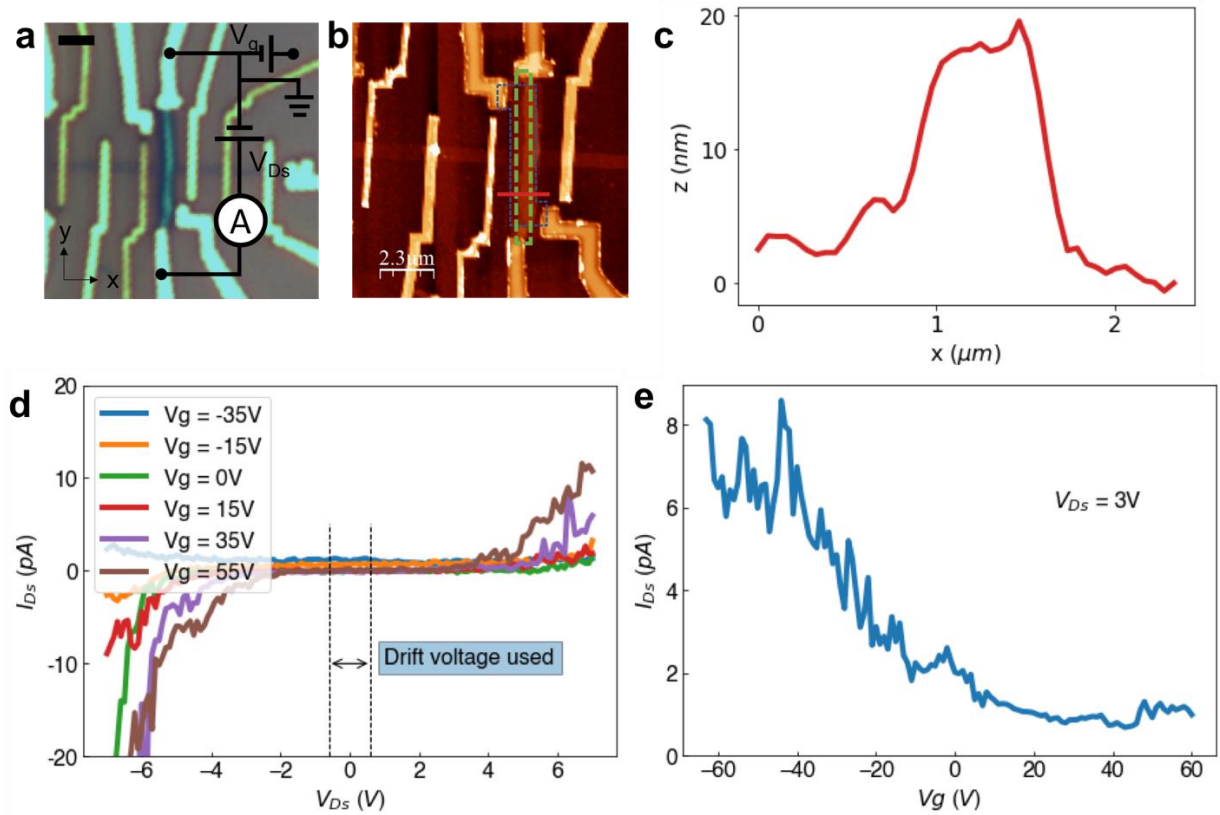

**Fig. S1 Electrical characterization of CGT.** (**a**) Optical image and (**b**) atomic force microscopy image of sample S1. The scale bar in panel **a** indicates 2 μm. The green and blue dashed line in panel **b** indicates the shape of the CGT flake and graphene, respectively. (**c**) The line-scan profile for the solid red line in panel **b**, indicating the CGT's thickness is ~20 nm. Two contacts to the CGT flakes were used to confirm the magnitude of the CGT resistance after the complete fabrication process. (**d**) The drift current $I_{Ds}$ versus drift voltage $V_{Ds}$ at different back gate voltage $V_g$. The n-doped Si substrate is used as the back gate. The $V_{Ds}$ used in our spin transport experiment stays between the dashed lines, where the CGT resistance is more than seven orders of magnitude larger than the two-terminal graphene resistance (14.5 kΩ). The low mobility of CGT[1] impedes effective electrical transport within it. Prior studies suggest that electrical detection is achievable at carrier densities with a range of at least $10^{14}$ cm$^{-2}$ (Refs. [2] and [3]). In contrast, the carrier density upper bound here is of $10^{12}$ cm$^{-2}$. (**e**) $I_{Ds}$ as a function of $V_g$ measured at $V_{Ds}$= 2 V and 300 K, indicating the CGT is a p-doped semiconductor.

**Note 2. Spin transport of pristine graphene.**

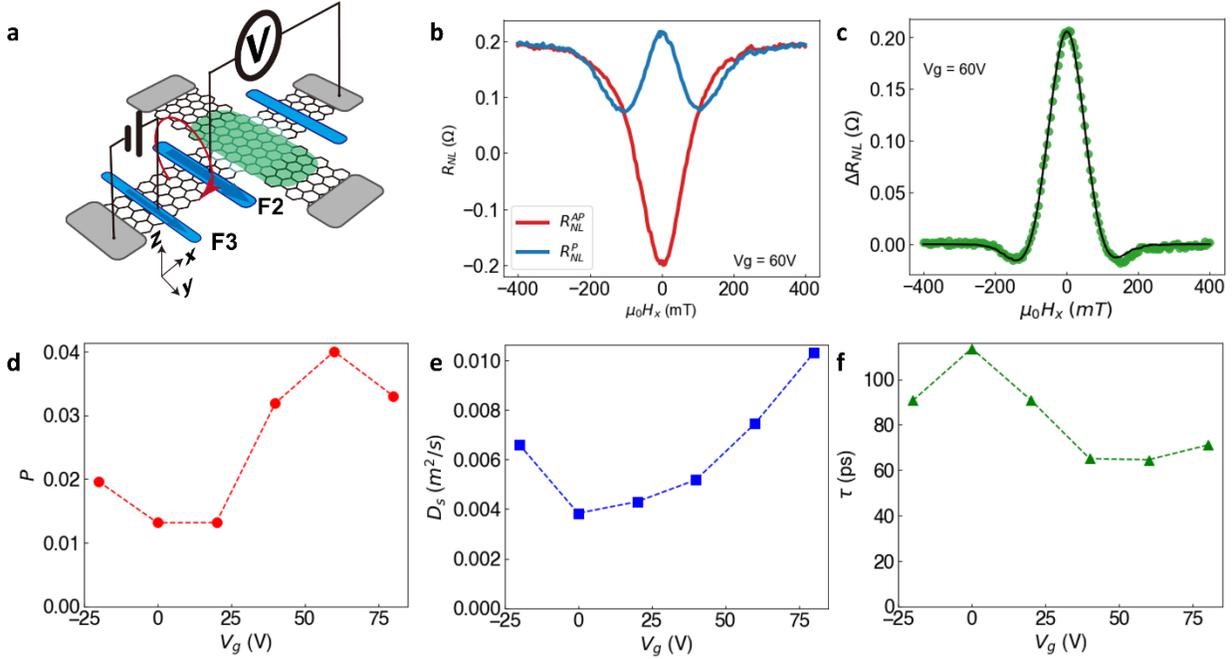

**Fig. S2 Symmetric Hanle precession arising from spin transport in pristine graphene.** (**a**) Measurement configuration for symmetric Hanle precession for pristine graphene with electrical spin injection and detection using two Co electrodes (F2 and F3). An in-plane magnetic field, $\mu_0 H_x$, is applied to induce precession of the *y*-polarized spins injected from the FM electrode into the graphene. (**b**) Non-local resistance as a function of $\mu_0 H_x$ measured at 100 K and $V_g = 60$ V using the configuration in panel **a**, with F1 and F2 electrodes set in a parallel ($R_{NL}^P$, blue line) and antiparallel ($R_{NL}^P$, red line) configuration. (**c**) Net symmetric Hanle precession signal extracted from the two curves in panel **b** by taking $\Delta R_{NL} = (R_{NL}^P - R_{NL}^{AP})/2$. The black solid line is a fit of the data to the solution of the Bloch equation (Eq. (6) in Note 10). Gate voltage dependence of (**d**) spin polarisation of Co/TiO$_x$ electrode, (**e**) spin diffusion constant, and (**f**) spin lifetime of pristine graphene. All data are taken from Sample 1 at 100 K.

**Note 3. Separating spin Hall effect from electrical spin injection when graphene is magnetic.**

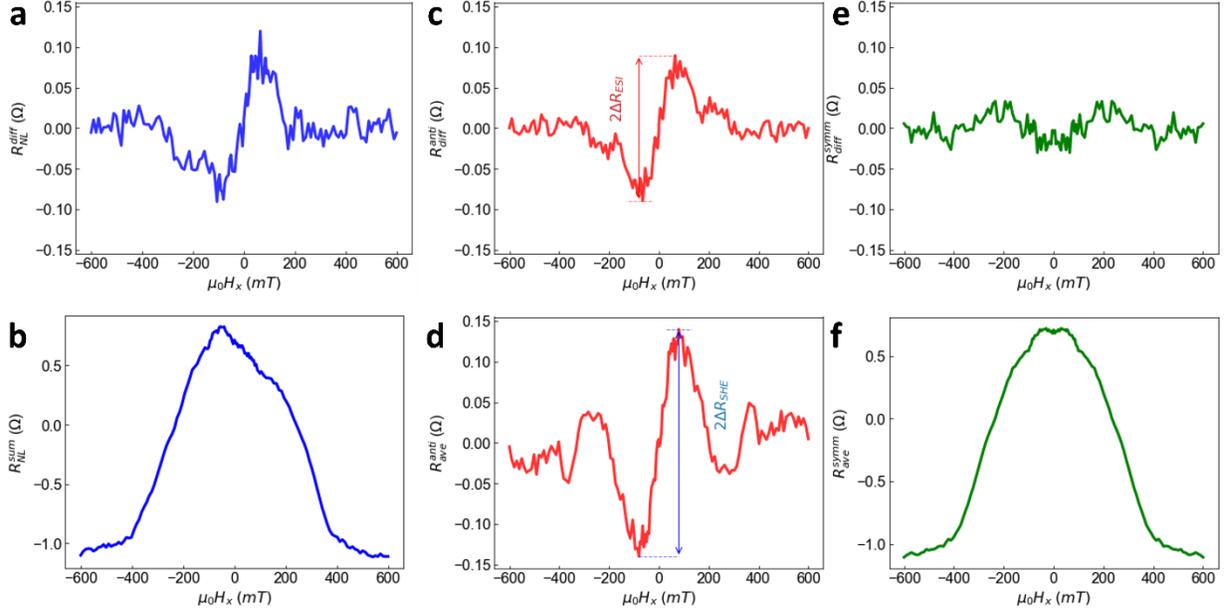

**Fig. S3 Separating SHE from ESI when graphene is magnetic.** The (**a**) difference $R_{NL}^{diff} = (R_{NL}^{\otimes} - R_{NL}^{\odot})/2$ and (**b**) average $R_{NL}^{ave} = (R_{NL}^{\otimes} + R_{NL}^{\odot})/2$ of the non-local resistance, with $R_{NL}^{\otimes}$ and $R_{NL}^{\odot}$ obtained when CGT is initialized along +z and −z, respectively, and measured while sweeping $\mu_0 H_x$ at 10 K and $V_g = 55$ V in Sample 1 (inset in Fig. 2d of the main text). The ESI-induced Hanle precession signal $R_{NL}^{diff}$, which depends on the initial CGT magnetization, is canceled in $R_{NL}^{ave}$, leaving any signal that is independent of the initial CGT magnetization. Here a linear background which might originate from the local ordinary Hall effect is removed in $R_{NL}^{ave}$. (**c, d**) Antisymmetric and (**e, f**) symmetric component of $R_{NL}^{diff}$ and $R_{NL}^{ave}$ is plotted, respectively. A clear Hanle precession signature from out-of-plane spins, as expected from SHE, is observed in panel **d**. The ESI-induced Hanle precession of panel **c**, which corresponds to Fig. 2d of the main text, shows a different amplitude but the same precession feature when compared to the SHE-induced Hanle precession of panel **d**. This is a remarkable result that confirms the spin origin of the two signals, because the precession of the spins generated from the two sources (SHE and ESI) in the same pristine graphene channel must be the same in both cases.

In the $R_{NL}^{ave}$ curve (Fig. S3b), in addition to the antisymmetric Hanle precession signal arising from SHE isolated in Fig. S3d), we can observe the existence of a symmetric background signal (Fig.S3f). The background signal exhibits a saturation field, which decreases with increasing the temperature, thus corresponding to the in-plane saturation field of CGT. A value of ~400 mT is observed at 10 K for such field. It is important to note that the antisymmetric signals occur well below the in-plane saturation field of CGT, indicating that they originate from spin precession. One possible origin of the background signal is from a local magnetoresistance of the magnetic graphene, where the stray current from the non-local measurement configuration provides a local resistance component. Another origin could be related to the magnetic multidomain state of CGT. The ESI efficiency between the magnetic multidomain state (at zero magnetic field) and the single magnetic domain state (when saturated along *x*), is different. In such a case, the spin current probed by spin precession would originate from the multidomain state, giving a lower output. However, at a high in-plane magnetic field, CGT is fully saturated along *x*, and exhibits the single domain state parallel to the FM electrode. In this case, the single domain spin polarisation ($P_{PGr}^{SD}$) would be 8.6 times larger than $P_{PGr}$ according to the difference between the spin precession and the baseline. With this model, the ESI efficiency for a single domain state would be estimated to be $P_{PGr}^{SD}$ = 124% ± 83%.

**Note 4. Finite element simulation of non-local spin signal.**

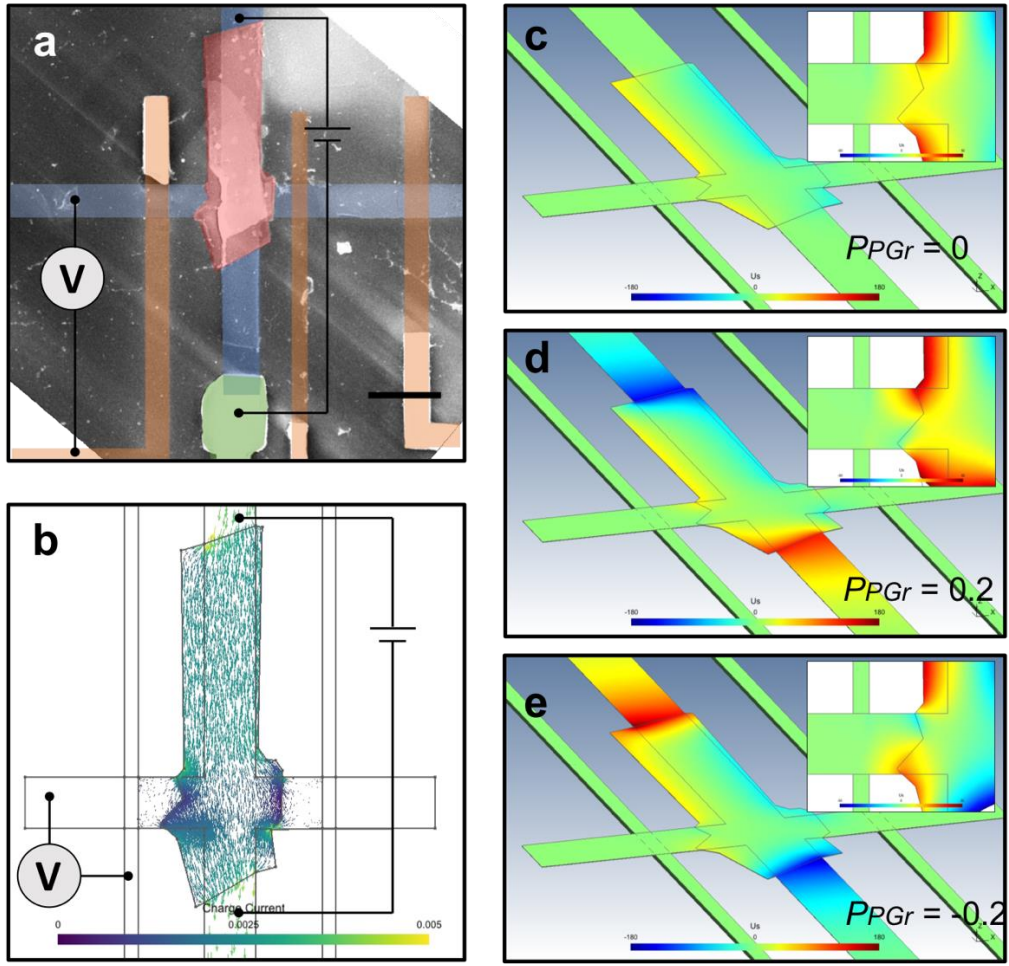

**Fig. S4.1 Finite element simulation of non-local signal including SHE and ESI of Sample 1.**
(**a**) False color SEM image of Sample 1, the scale bar is 2 μm. (**b**) Finite element simulation result of the charge current distribution (normalized unit). The charge-to-spin measurement configuration is also illustrated. The mesh size of the simulation is set to be 5 nm. (**c**) The spin chemical potential mapping at $\theta_{SH}$= 5%. The inset is a zoom for the cross-section region close to the detecting FM electrode. A clear spin accumulation at the edges transverse to the current direction can be observed, originated from the SHE. We switch on the magnetism of PGr by adding a polarisation of (**d**) $P_{PGr}$ = 0.2 and (**e**) $P_{PGr}$= −0.2. A clear spin accumulation, originated from ESI, can be observed at the boundary between the pristine graphene and the proximitized graphene at the current path. The result shows that, independent to the FM electrode, for achieving the same non-local spin signal, the polarisation $P_{PGr}$ should be 9.7 times larger than the spin Hall angle $\theta_{SH}$, which is because the geometry of the device is not optimal for ESI.

The parameters of the proximitized graphene below $T_c$ are fitted as follows as shown in Fig. S4.1. The spin diffusion length of pristine graphene ($\lambda_s^{gr} = \sqrt{\tau_s^{gr} D_s}$ =650 nm) and the Co polarisation ($P_{Co}$= 3.8%) were deduced from the Hanle spin precession of two Co electrodes (Note 2). Then, these parameters were fixed, albeit with associated errors, for the fitting of the antisymmetric Hanle curve (Fig. S3d) originated from the SHE with the numerical model described in Note 10, obtaining a spin Hall angle $\theta_{SH}$ =2.4 ± 1.6 %, and a spin diffusion length $\lambda_s^{PGr}$= 250 nm ± 890 nm. After that, all the parameters with associated errors are then incorporated into a 3D simulation, leaving the spin polarisation of the proximitized graphene ($P_{PGr}$) as the only unknown parameter. This simulation was used to fit $P_{PGr}$ (14.5% ± 9.6 %). The propagation error when fitting in this manner results in a larger error bar.

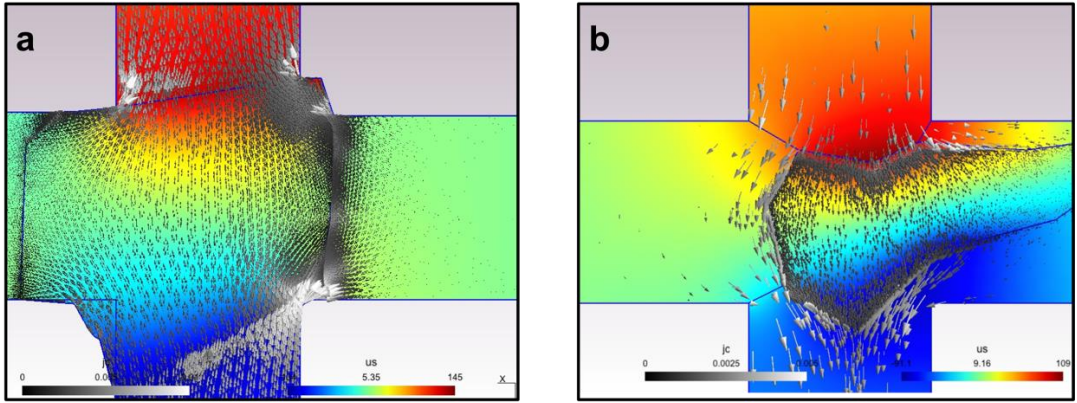

**Fig. S4.2 Finite element simulation of non-local signal of Sample 2.** (**a**) The sketch shows the results of a simulation where a charge current is applied through the Hall arm from N3 to N4 in the PGr1 region shown in Fig. 3b of the main text. The spin chemical potential mapping is shown in rainbow colors, and the charge current mapping is shown in gray scale arrows. The simulation reveals a strong spin accumulation at the boundary between the PGr and pristine graphene. In (**b**), the current is applied through the Hall arm from N1 to N2, and the spin chemical potential mapping is shown.

The spin polarisation of the proximitized graphene (PGr) in Sample 2 was determined as follows. We used the sheet resistances of the PGr (2.3 kΩ) and the pristine graphene (0.8 kΩ) at a gate voltage of 50 V, as shown in Fig. S5.2, to fit the symmetric Hanle spin precession data in Fig. 3e using Eq. (6) in Note 10. This fitting yielded a spin diffusion length of the pristine graphene, $\lambda_s^{gr}$ = 1240 ± 170 nm, and an apparent spin injection efficiency of 1.8% ± 0.3%. Subsequently, we incorporated and fixed this spin diffusion length, along with the spin Hall angle $\theta_{SH}$ =2.4% and the spin diffusion length $\lambda_s^{PGr}$= 250 nm of the proximitized graphene obtained from Sample 1, into a 3D simulation, where $P_{PGr}$ is the only fitting parameter. This approach resulted in an estimated spin polarisation of $P_{PGr}$ = 61.5 ± 9.5%. Here, we assumed the spin polarisation of the two proximitized graphene regions to be the same for simplicity.

**Note 5. Gate-tunable charge transport properties and charge-to-spin conversion.**

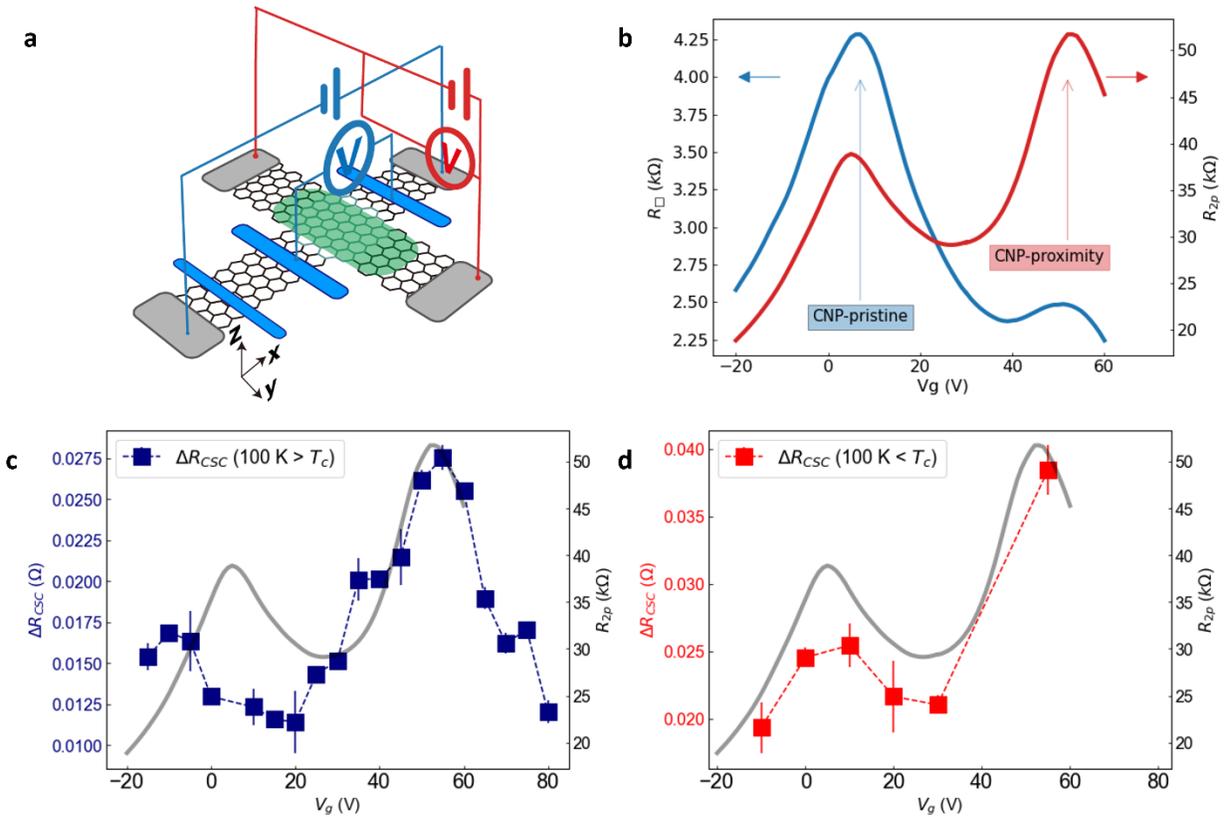

**Fig. S5.1 Gate-tunable charge transport properties and charge-to-spin conversion of Sample 1.** (**a**) Measurement configuration for the charge transport properties of graphene. Square resistance of the cross-section region graphene (4-point resistance, in blue) and at the cross-junction of the Hall bar (2-point resistance, in red). (**b**) Square resistance of the cross-section graphene (blue solid line) and 2-point resistance of the CGT-covered graphene at the cross-junction of the Hall bar (red solid line) measured at 100 K. The CNPs for the pristine graphene is around 8 V, while the proximity graphene is around 55 V. (**c**) Amplitude of the antisymmetric Hanle precession signals $\Delta R_{CSC}$, measured at 100 K as a function of $V_g$. The charge-to-spin conversion shows a peak at the CNP of the proximitized graphene. (**d**) $\Delta R_{CSC}$ as a function of $V_g$ measured at 40 K, by setting the magnetization always pointing upwards. The maximum output is also observed at the CNP of the proximitized graphene. Data are presented as the half-amplitude at the maximum minus the minimum of the antisymmetric curve with the error bars are calculated using the standard deviation of the antisymmetric component of the Hanle precession signal within the saturation region at the highest applied fields, where the signal should be constant and allows for a consistent quantification of the noise in the measurement. Here, $\Delta R_{CSC}$ includes the charge-to-spin conversion originating from both SHE and the ESI. All data correspond to Sample 1. In the present experiment, only back gate is used, which tunes both the carrier concentration and the vertical electric field in the heterostructure. Although the two effects cannot be disentangled

without a top gate, the range of voltages used corresponds to small electric fields and we only expect to modulate a carrier concentration in the graphene[4].

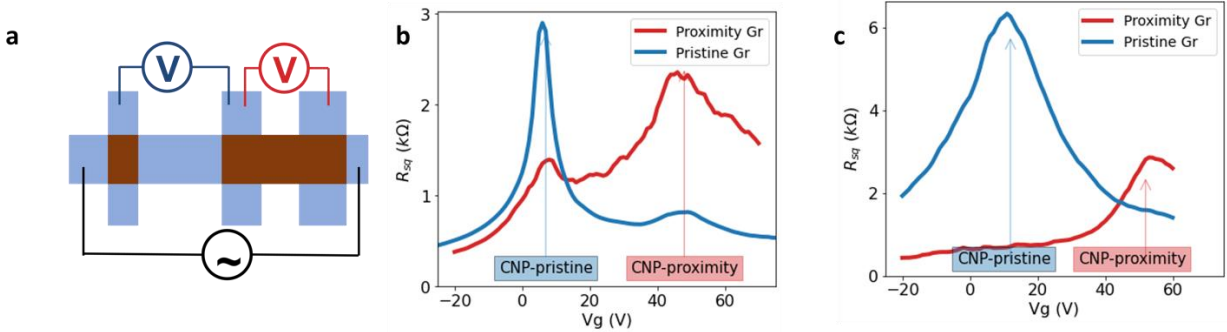

**Fig. S5.2 Gate-tunable charge transport properties of Sample 2.** (**a**) Measurement configurations of the longitudinal resistance of the pristine graphene (blue circuit) and the proximitized graphene (red circuit). (**b,c**) Square resistance ($R_{sq}$) as a function of back-gate voltage ($V_g$) for Sample 2 measured at 50 K (**b**) and Sample 3 measured at 10 K (**c**), using configuration in panel **a**. The CNP of the proximitized graphene is approximately 55 V for both samples, indicating a similar p-type doping by the graphene-CGT interface. This consistent doping level suggests that the interface remains stable throughout the fabrication processes used for each sample. The charge transfer between CGT and graphene, which has also been observed in previous publications[5–7], possibly arises from the presence of a vertical electric field within the heterostructure.

**Note 6. Atomic force microscopy of CGT/graphene heterostructures.**

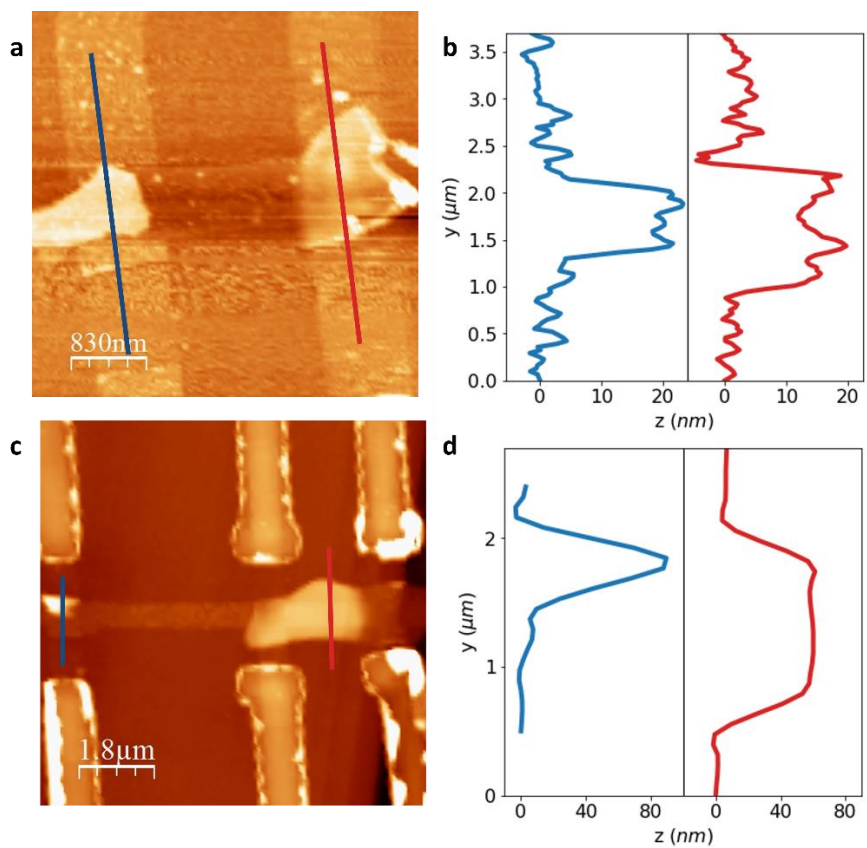

**Fig. S6 Atomic force microscopy characterization**. Result for (**a**) Sample 2 and (**c**) Sample 3. The line scan profiles of the heterostructure area are shown in (**b**) and (**d**), with the blue/red curves representing the left side / right side CGT flakes, respectively. This thickness of the CGT flake is 20 nm and 15 nm in Sample 2, and 85 nm and 60 nm in Sample 3.

**Note 7. Raman spectra of CGT/graphene heterostructures.**

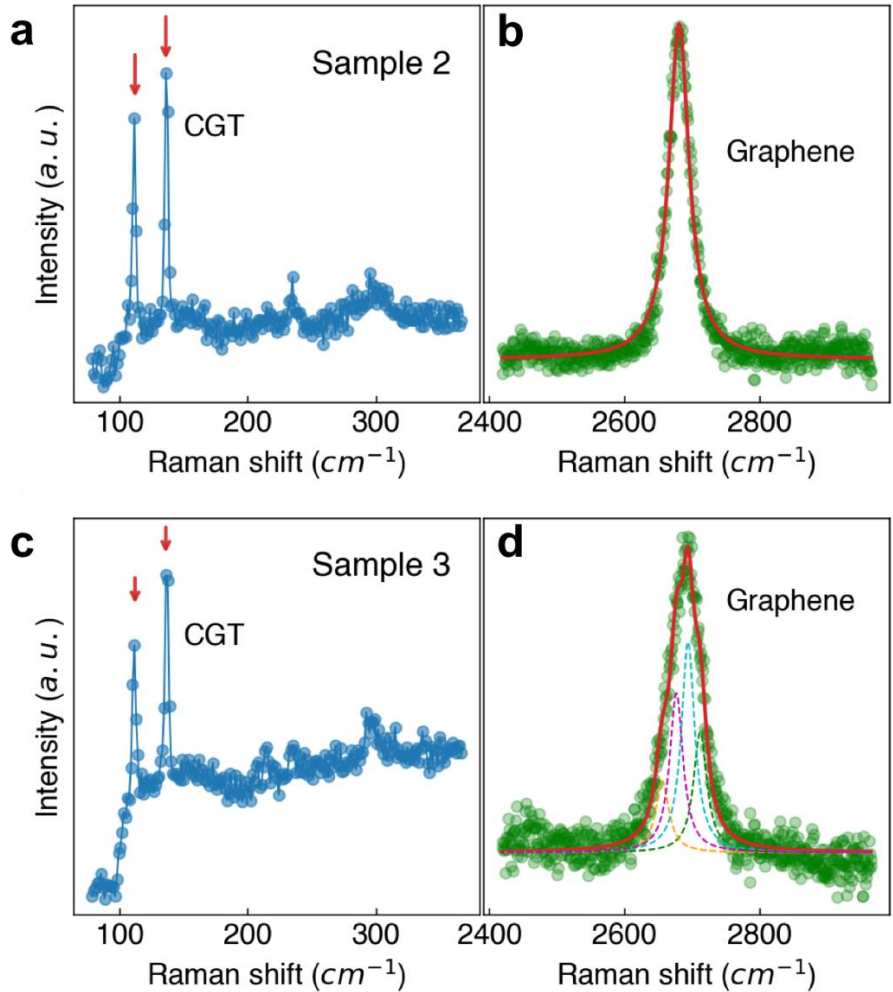

**Fig. S7 Raman spectra of CGT/graphene heterostructures.** Representative micro-Raman spectra of the CGT flakes corresponding to Sample 2 and Sample 3 (**a**), (**c**), and the graphene channel (**b**), (**d**) after the electrical measurements using a green polarized laser (532 nm). All the spectra are taken at 300 K. The two well-separated peaks in panels **a** and **c** at 110 cm$^{-1}$, and 137 cm$^{-1}$ ascribed to the E2g and A1g symmetry Raman modes, indicating the original structure with a low oxidation condition[8] after all the fabrication and measurement process. The 2D peak of graphene in panel **b** is fitted with one Lorentzian function, indicating single layer. The 2D peak of graphene in panel **d** is fitted with four Lorentzian functions with Full Width Half Maximum of 24 cm$^{-1}$ (Ref. [9]), therefore, corresponding to a thickness of two layers.

**Note 8. Extra measurement of the anomalous Hall effect.**

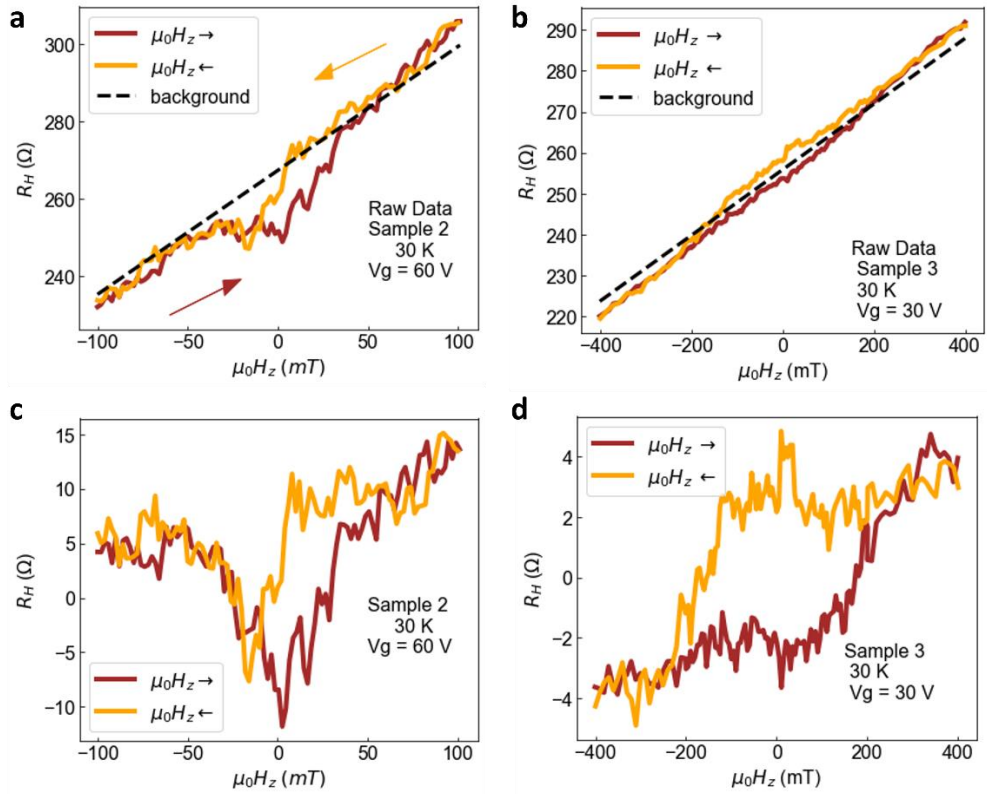

**Fig. S8 Anomalous Hall effect.** Raw data of the transverse Hall resistance of (**a**) Sample 2 measured at 30 K and $V_g$= 60 V, and (**b**) Sample 3 measured at 30 K and $V_g$= 60 V, showing a clear anomalous Hall effect with a hysteresis feature even without the removal of any background. The linear baseline originating from the ordinary Hall effect is plotted in black dashed lines. (**c**), (**d**) Transverse Hall resistance after subtracting the linear background shown in (**a**) and (**b**), respectively.

**Note 9. The coexistence of spin Hall effect and the electrical spin injection in an all-2D lateral spin valve.**

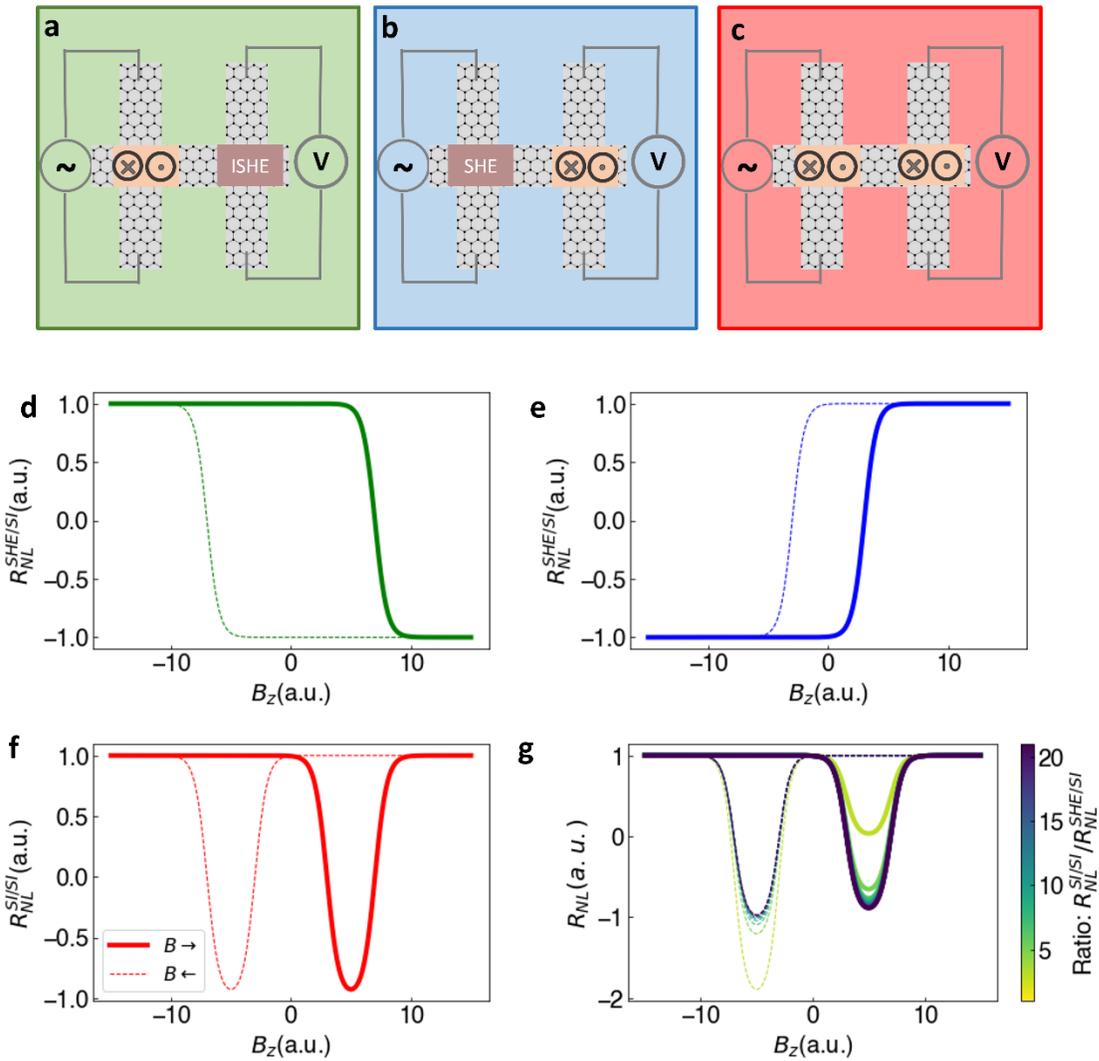

**Fig. S9 The coexistence of spin Hall effect and the electrical spin injection.** A schematic representation of the three possible spin-charge interconversion scenarios in the all-2D lateral spin valve. The original mechanisms for spin generation/detection are: (**a**) electrical spin injection/inverse spin Hall effect, (**b**) spin Hall effect/electrical spin detection, and (**c**) electrical spin injection/electrical spin detection for the two proximitized regions, respectively. The corresponding non-local resistances as a function of $B_z$ are illustrated in (**d**), (**e**), and (**f**). The amplitude is set arbitrarily, and a sigmoid function is introduced to indicate the multidomain state of the CGT. It should be noted that the spin Hall effect/inverse spin Hall effect, which would contribute as a constant background to the non-local signal, is not depicted. The final non-local spin signal, shown in (**g**), is a combination of all factors. A color gradient is employed to indicate the varying contributions from spin Hall effect and the electrical spin injection. As spin injection becomes more prominent, the imbalance in the non-local resistance becomes increasingly apparent and distinct.

A fourth possible contribution could arise from the anomalous Hall effect present in the proximitized region. The anomalous Hall effect induces a transverse charge current, which changes sign with the magnetization. For an ideal geometry, it is not possible to non-locally pick up the voltage associated with this transverse current. However, due to the asymmetry of the device geometry, it is possible to detect such a non-local voltage in the real device, something that has been studied with finite element simulations in one of our previous papers[10]. This non-local resistance would have the same shape as the signal shown in panels **d** and **e** and, therefore, will be indistinguishable from the electrical spin injection/inverse spin Hall effect and the spin Hall effect/electrical spin detection contributions.

## Note 10. Spin precession fitting process.

The spin propagation in our devices using the Bloch equations:

$$D_S \nabla^2 \vec{\mu} - \frac{\vec{\mu}}{\tau_s} + \vec{\omega} \times \vec{\mu} = 0 \tag{1}$$

where $\vec{\mu}$ is the spin accumulation, $D_S$ the spin diffusion constant, and $\tau_s$ the spin lifetime. $\vec{\omega} = g\mu_B \vec{B}$ is the Larmor frequency, $g = 2$ is the Landé factor, $\mu_B$ is the Bohr magneton, and $\vec{B}$ is the applied magnetic field.

The spin accumulation at the detection position $x = x_{det}$ is then converted into a voltage with

$$V_{det} = P_{FM} \frac{\mu_y(x_{det})}{e} \tag{2}$$

$P_{FM}$ is the polarisation of the ferromagnetic electrode. The voltage is usually normalized by the charge current $I_c$, giving the non-local resistance

$$R^s = \frac{V^s}{I_c}. \tag{3}$$

In the case of the contact resistance between Co and graphene being large, Spin dispassion behavior on the magnetic contact can be neglected.

The non-local resistance is also affected by the contact pulling effect[11], since the magnetic field Bx is applied perpendicular to the easy axis of the ferromagnets. The magnetization is then pulled by an angle in the field, giving the non-local resistance an additional term. The measured non-local resistance is

$$R_{NL} = \pm R^s \cos(\beta_1)\cos(\beta_2) + R^{s0}\sin(\beta_1)\sin(\beta_2). \tag{4}$$

Here $\pm$ represents the non-local signal with parallel (P) and antiparallel (AP) configuration. $\beta_1$ and $\beta_2$ correspond to the two magnetic electrodes used as injector and detector. $R^{s0}$ is a constant that corresponds to the spin signal at zero magnetic field. The net spin signal $\Delta R_{NL}$ is then expressed as:

$$\Delta R_{NL} = \frac{R_{NL}^P - R_{NL}^{AP}}{2} = R^s \cos^2(\beta). \tag{5}$$

Because of the opposite spin precession sign of $R_{NL}^P$ and $R_{NL}^{AP}$, the average $R_{ave} = (R_{NL}^P + R_{NL}^{AP})/2$ is then is proportional to $sin(\beta_1)sin(\beta_2)$. For simplicity, we consider that the two FM electrodes have the same pulling effect with the angle, $\beta = \beta_1 = \beta_2$.

We thus use the following equations for the fitting of the spin precession measurements in the main text:

### (i) The symmetric Hanle precession experiment:
For the Hanle precession with two Co electrodes or with two proximitized graphene regions, the spin transport parameters are obtained with

$$\Delta R_{NL} = \frac{P_{FM}^2 \cos^2(\beta) R_{sq} \lambda_s}{W_{gr}} \Re \left\{ \frac{e^{-\frac{L}{\lambda_s}\sqrt{1-i\omega\tau_s}}}{\sqrt{1-i\omega\tau_s}} \right\}, \qquad (6)$$

where $\lambda_s$ is the spin diffusion length, $\tau$ is the spin lifetime, $W_{gr}$ is the width of the channel, $R_{sq}$ is the square resistance, and $L$ indicates the distance between the two ferromagnetic electrodes.

### (ii) For the antisymmetric Hanle precession of the SHE measurement:
The non-local resistance $\Delta R_{SHE}$ is given by

$$\Delta R_{SHE} = \frac{\theta_{SH} R_{sq} \overline{I_{Sz}}}{I_c} \cdot \cos(\beta) \qquad (7)$$

where $\overline{I_{Sz}}$ is the average spin current in the cross-junction area of the Hall bar, which is defined as $\overline{I_{Sz}} = \frac{1}{W_{cr}} \int_L^{L+W_{cr}} I_{Sz}(x) dx$, where $L$ is the distance from F1 electrode to the edge of the Hall cross and $W_{cr}$ is the width of the cross. To determine $I_{Sz}(x)$, we use a numerical model, using the following boundary conditions:

1. Spin accumulation μ is continuous.
2. Spin current is continuous except at F1 with spin injection by the charge current, by $\Delta I_s = I_c \cdot P_{FM}/2$;
3. Spin current is zero at the sample ends.

We write these boundary conditions in a matrix $X$ which fulfills: $MX = Y$, where $M$ contains the coefficients $M = A, B, C \ldots$ and $Y = (0, \ldots, \frac{I_c P_{FM}}{2}, 0, \ldots)$ and use the Moore-Penrose inverse to invert $X^{-1}$ and obtain $M = YX^{-1}$. Using the numeric solution described above, we input the measured net spin signal $\Delta R_{NL}$, square resistance $R_{sq}$, pulling factor $\cos^2(\beta)$, channel size $W_{gr}$ determined by SEM. Finally, we assume the diffusion coefficients for the proximitized region and the pristine region to be equal to reduce the fitting parameters.

**Note 11. Mechanisms of the observed spin Hall effect.**

The SHE can arise from different mechanisms (intrinsic, skew scattering, side jump) and it is not easy to separate them experimentally in graphene-based heterostructures[12,13]. The intrinsic mechanism of SHE in TMD-proximitized graphene was shown to arise mostly from the valley-Zeeman SOC, see prediction[14,15] by Garcia et al.. However, valley-Zeeman SOC alone does not lead to a SHE, and only in combination with Rashba and/or intrinsic SOC is the spin Hall conductivity finite.

Valley-Zeeman SOC is also present in CGT/graphene heterostructures, as evidenced by the anisotropic spin relaxation reported[7] by Karpiak et al., akin to the same phenomenon observed in TMD/graphene[12,15]. By fitting the low-energy model of graphene bands to the density functional theory calculations[7] of CGT/graphene with SOC, values of the valley-Zeeman SOC, $\lambda_{VZ} = 0.113$ meV, and the Rashba SOC, $\lambda_R = 0.253$ meV. However, a calculation of the intrinsic SHE from these values has not been performed yet. As in the case of TMD/graphene, a univocal relation between the spin Hall angle and the valley-Zeeman SOC is not expected.

The SOC energy values are notably lower than those found in a TMD/graphene heterostructure, such as $WS_2$/graphene, which has a valley-Zeeman SOC of $\lambda_{VZ} = 1.12$ meV and a Rashba SOC of $\lambda_R = 0.36$ meV, as reported[16] by Gmitra et al.. The co-existence of both intrinsic and extrinsic mechanisms in TMD/graphene heterostructures has been discussed in the previous papers ($MoS_2$/graphene[17], $WS_2$/graphene[12], $WSe_2$/graphene[13]). With a similar structure, the presence of an extrinsic mechanism contributing to the SHE in our heterostructure is also possible. Theoretical calculations[18] indicate that extrinsic mechanisms, such as skew scattering in 2D Dirac fermions, can be influenced by the spin texture of the energy bands, leading to a SHE contribution.